\documentstyle[aps,multicol,epsf]{revtex}
\begin{document}
\draft

\title{Systems with Multiplicative Noise: Critical Behavior from
KPZ Equation and Numerics}

\author{Yuhai Tu, G. Grinstein, and M.A. Mu{\~n}oz}

\address{IBM Research Division, T.J.
Watson Research Center, P.O. Box 218, Yorktown Heights, NY 10598}
\date{\today}
\maketitle

\begin{abstract}We show that certain critical exponents of
systems with multiplicative noise can be obtained from exponents of the
KPZ equation.  Numerical simulations in 1d confirm this prediction, and
yield other exponents of the multiplicative noise problem.  The numerics
also verify an
earlier prediction of the
divergence of the susceptibility over an entire
range of control parameter values, and show that the exponent governing
the divergence in this range varies continuously with control parameter.
\end{abstract}

\begin{multicols}{2}
\narrowtext

Langevin equations --
first order partial
differential equations in time
containing Gaussian random noise terms -- capture the macroscopic
physics of many classical, stochastic, many-body systems\cite{VK}.
In the most common situation, one that includes much
of equilibrium statistical mechanics, the
noise amplitude is simply a constant.  There are however, important classes of
problems in which the noise amplitude is proportional to a positive power,
$\alpha$, of the field variable itself.  The well-known
case $\alpha =1/2$ describes
the physics of directed percolation and its many and diverse
realizations\cite{DP}
in, e.g., models of catalysis, the spread of epidemics or forest fires,
and propagating star formation.  Other
systems
in which the dominant source of noise is
external,
among them certain chemical reactions
and problems in quantum optics,
are characterized by $\alpha=1$\cite{Brand}.  These so-called
``multiplicative-noise" systems (MNS)\cite{general}
are the subject of this paper.

The generic problem with a single-component field in this class is described by
the Langevin equation

\begin{equation}
\partial n(\vec x ,t) /\partial t = \nabla^2 n - rn -un^{2+\rho} + n \eta ~~,
\label{MNS}
\end{equation}
where $r$ and $u$ are parameters, and $\eta$ is a Gaussian noise variable
with correlations $<\eta(\vec x ,t) \eta(\vec x' ,t')> = D \delta (\vec x -
\vec x') \delta (t-t')$, for some noise strength $D$.
We consider both quadratic and cubic nonlinearities, $\rho =0$ and $\rho
=1$, respectively.

In a previous paper\cite{legend}, we analyzed the phase structure of this
model, which has two phases, an ``active" phase with $<n> ~ > 0$,
and an ``absorbing" phase with $n(\vec x) =0 $ for all $\vec x$,
that occurs for sufficiently large $r$.  These phases are
separated by a critical point at $r \equiv r_c $, where
$r_c =0$ in mean-field
theory.  (Note that the vanishing of the noise amplitude with $n$
makes the right side of Eq. (\ref{MNS}) vanish when $n(\vec x)$ vanishes,
thereby causing all dynamics to cease and making the absorbing configuration
$n(\vec x)=0$ a potentially stable phase.)  We showed that model (\ref{MNS})
had a
critical
dimension $d_c =2$, below which the transition is governed by an analytically
inaccessible strong-coupling
fixed point.
For $d>2$, on the other hand, the transition is
governed by the ``weak-coupling" Gaussian fixed point
with mean-field exponents for $D$ less than a critical
value $D_c$.  For $D>D_c$, however, the strong-coupling
fixed point is the stable one.  A multicritical point
occurs at $D=D_c$.

In this paper we argue that the critical behavior
of MNS
is actually governed by
the fixed point of the Kardar-Parisi-Zhang
(KPZ) model of growing interfaces\cite{kpz}.  This mapping
is consistent with the phase diagram proposed in ref.
\cite{legend}, and with known exponents of the
weak-coupling and multicritical fixed points.  It also
allows us to express
the dynamical exponent $z$ and the
correlation length exponent $\nu$
of the strong-coupling transition
in terms
of the KPZ exponents.
These exponents are
found to be
independent of the degree, $2+\rho$, of the nonlinearity.
A lower bound of unity for the order parameter exponent
$\beta$ in the case $\rho = 0$
also emerges.  We confirm these predictions by
calculating numerically in 1d the four independent exponents
%which scaling
%arguments predict are required to
characterizing the strong-coupling transition.
We also confirm
the rather striking
prediction in ref. \cite{legend} of an
entire domain
of $r$ values, encompassing
the critical value $r_c$, in which the susceptibility of the
system diverges.  The numerics show that, like in the
exactly solvable single variable (0d) problem, this region of
infinite susceptibility extends to both sides of the critical
point, and is controlled by a fixed line with continuously
varying critical exponents\cite{legend}.
%Finally, we verify numerically the
%scaling predictions of ref. \cite{legend}.

To establish the connection between model (1) and the KPZ theory,
note that
the field $n(\vec x ,t)$ in (1)
will remain positive if $n(\vec x ,0) > 0$ for all $x$.  In this case one can
perform the Hopf-Cole\cite{HC}
change of variable $n(\vec x ,t) = e^{h(\vec x ,t)}$,
producing the equation\cite{Pik}

\begin{equation}
\partial h /\partial t = -r + \nabla^2 h +(\nabla h)^2 -
u e^{(1+\rho) h} + \eta ~~.
\end{equation}

Aside from the $u$ term this is precisely the KPZ equation
(wherein the standard KPZ nonlinearity $(\nabla h)^2$
has coefficient unity).
Note however that either in
the absorbing phase or at the critical point, the steady-state value
of $n$ is zero, whereupon the steady-state value of $h$ is
$- \infty$.  Thus the u term vanishes in steady state,
leaving one with precisely the KPZ theory.

Recall\cite{kpz}
that the phase diagram of the KPZ equation consists
of a unique strong coupling phase for $d \leq d_c=2$, and both
weak and strong coupling
phases separated by a multicritical point for $d>2$.
The weak and strong-coupling regimes
occur for noise strengths $D$
that are respectively smaller and larger than a critical value.
This phase diagram is
thus encouragingly similar to that of the MNS.
Also,
the dynamical critical exponent
$z$ is 2 for both MNS\cite{legend} and KPZ\cite{Magic}
along the line of
weak-coupling transitions and at the multicritical point,
for $d>2$.

Let us now consider the critical exponents for the strong-coupling
transition.
The dynamical
critical exponent $z$ for the MNS
can be computed from the steady-state
behavior of response
functions right at the critical point, and hence
is identical to the value of $z$ in the KPZ theory.  In particular then,
$z=3/2$ for $d=1$\cite{kpz}.

To argue that other exponents of the MNS
can also be obtained from
the KPZ equation requires us to consider the active state of model (\ref{MNS}).
For $r$ slightly less than $r_c$, $0 < ~~~ <n> ~ \ll 1$ in steady state,
implying that $h_0 \equiv <h>$
is very large and negative.  Writing $h(\vec x,t)
\equiv <h> + \delta h(\vec x, t)$, one obtains an equation for
$\delta h$ that is identical to the KPZ equation except for the extra
nonlinear
term $-u' e^{(1+\rho) \delta h}$, where $u' \equiv u e^{(1+\rho) h_0}$.
Noting that the leading
nontrivial term in the expansion of this nonlinearity in powers of $\delta h$
is the linear ``mass" term $-u' (1+\rho) \delta h$, we infer that the main
effect of $-u' e^{(1+\rho) \delta h}$ is to produce a finite correlation
length $\xi$ at which the power law correlations of the KPZ equation
are cut off and replaced by exponential behavior.  One concludes that $\xi$
must be the correlation length of the corresponding MNS; its
divergence as $r \rightarrow r_c$ is governed by the critical exponent
$\nu$.  In the critical region, i.e., on length scales
$ | \vec x | << \xi$, the $u'$ term is negligible, so critical correlations
can be computed from the KPZ equation.

To calculate $\nu$ from KPZ, take the expectation value of the $\delta h$
equation, recall that $<\delta h> =0$, and write the extra nonlinear
term as $-u<n^{1+\rho}>$, to obtain

\begin{equation}
-r + <(\nabla h)^2> -u <n^{1+\rho}> = 0 ~~.
\label{nu}
\end{equation}

At the critical point, $r=r_c$ and $n=0$, so $-r_c
+ <(\nabla h)^2>_c = 0 ~~$.  Subtracting this equation
from (\ref{nu}) yields
$ \delta W = -\delta r + u <n^{1+\rho}>$,
where
$\delta r \equiv r_c -r$, and
$\delta W \equiv <(\nabla h)^2> - <(\nabla h)^2>_c$.
>From the standard scaling of the KPZ equation\cite{kpz},
one has $\delta W \sim - C \xi^{2(\chi -1)}$, where $C$ is a
positive constant, and $\chi$ the roughness exponent of a
KPZ interface; i.e., $ <(h(\vec x ,t) - h(\vec 0 ,t))^2> \sim
x^{2 \chi} $ for KPZ in steady-state.
Since $C > 0$,
$ \delta r > 0$  in the
active phase, and $<n^{1+\rho} > \sim (\delta r)^{\beta_{1+\rho}} > 0$,
the equation for $\xi$ has a solution for small $\delta r$
only if the exponent $\beta_{1+\rho}$ is greater than unity.
For the quadratic nonlinearity,
$\rho =0$, so $<n^{1+\rho}>$ is the order parameter $<n>$, whereupon
this constraint places the nontrivial bound $\beta >1$ on the
order parameter exponent
$\beta \equiv \beta_1$.  For any value of $\rho$, one then
obtains the result $\xi\sim \delta r^{-\nu}$ with $\nu = 1/(2-2 \chi)$.  (Note that in cases where the
KPZ interface is smooth, $\chi = 0$ in this formula.)

%In our previous study of the critical behavior of the multiplicative
%noise system (MNS), we have found from RG analysis that in dimensions equal
%to or lower
%than 2 or for sufficiently large noise in higher dimensions, the transition
%from active to absorbing state is controlled by a strong coupling fixed
%point and is therefore beyond the reach of any perturbative analysis.
%To study the
%property of the strong coupling fixed point, we have to resort to
%non-perturbative method or numerical study.

We now describe the numerical simulation of (\ref{MNS}) in 1d.
We discretize the continuum equation as:

\begin{eqnarray}
n_{i}(t+\Delta t)&=&n_{i}(t)+\Delta t [-r n_{i}(t)-u{n_{i}(t)}^{2+\rho}\nonumber\\
&+&(1/\Delta x^2)
(n_{i+1}(t)+n_{i-1}(t)-2n_{i}(t))]\nonumber\\&+&\sqrt{D\Delta t}
n_{i}(t)\eta_{i}(t)
\;\;\;\;\; (i=1,2,\cdots,N),
\label{discr}
\end{eqnarray}
where $\Delta x$ is the lattice spacing, $\Delta t$ is the
time step and $\eta_{i}(t)$ is a Gaussian random number with unit standard
deviation. By choosing the argument of the field multiplying the noise
as in Eq. (\ref{discr}),
we are using the Ito interpretation\cite{VK}
of the stochastic differential equation.
We set $\Delta x=1$ , $\Delta t=0.02$, and use periodic boundary
conditions with the system size $L=N\Delta x$.
We also
fix the noise amplitude $\sqrt{D}=4$, and vary the linear
coefficient $r$ as the only control parameter. In the following,
we present our numerical results for $\rho=1$\cite{rho}.

In numerical simulation, due to the finite time step $\Delta t$,
the property of (1) that $n(x ,t)>0$ if $n(x ,0) >0$ for all $x$
is lost. However, it is
easy to fix this problem by setting $n_{i}(t)$ to zero
if its value becomes negative under Eq. (\ref{discr}).  The effect of this
modification is easily estimated. For a single time
step, we can neglect the second term on the RHS of (\ref{discr}) because it is
of higher order in $\Delta t$ than the noise term. Then
setting negative values of $n_i (t)$ to 0 is equivalent to truncating
the probability distribution of $\eta_{i}(t)$ so that its minimum is
set by $\eta_{min}=-1/\sqrt{D\Delta t}$, and replacing all the
$\eta < \eta_{min}$ by $\eta_{min}$.  This means, however, that the mean of
$\eta_{i}(t)$ is
no longer zero, and thus a deterministic term proportional to $n_{i}(t)$
is generated.  The resulting
effective linear coefficient $r_{eff}$ can be roughly
estimated as:
$r_{eff}=r-\sqrt{D/2\pi\Delta t}\int_{-\infty}^{\eta_{min}}
(\eta_{min}-\eta)\exp (-\eta^2 /2) d\eta$.
The effective strength of the noise is also changed because of the truncation.
These changes in parameters should not, however, alter the universality
class of the transition\cite{RFT}.

The first step in studying the critical behavior of MNS numerically
is to locate
the critical point.
Starting with uniform initial conditions, and letting the equation evolve
long enough to reach steady state, we compute
$M=<n_{i}(t)>$ for different values of $r$, where $<~~>$
denotes both spatial and
temporal averages, as well as averages over different independent runs.
We average over between
$2*10^5$ and $6*10^6$ time steps, and up to 100 independent
runs,
depending on the system size.
To handle finite-size effects, we studied different system sizes:
$N=100,200,400,1000$.  In fig. 1(a),
we show the dependence on the inverse
system size $1/N$
of the critical point, $r_c (N)$, defined by M first becoming
numerically indistinguishable from zero in every run.
Extrapolating the fitted line\cite{inverse}
to $N=\infty$
determines a critical value $r_c \approx
-2.18$.
Fig. 1(b) shows M vs. $\delta r \equiv r_c-r$
on a log-log plot for N=400.  The best fit to
$M \sim (\delta r)^\beta$ yields\cite{discrete} $\beta=1.70\pm 0.05$.

Also depicted in fig. 1(b) are the higher order moments of the $n$ field:
$M_{m}=<n^{m}>$, with $m=2,3$. Evidently, $M_{2}$ and $M_{3}$ also
have a power law
dependence on $\delta r$: $M_{m}\sim (\delta r)^{\beta_{m}}$ with
$\beta_{2}$ and $\beta_{3}$ being equal to $\beta$ within our numerical
accuracy.
It is not surprising that there is anomalous scaling, i.e.,
$\beta_2 \neq 2 \beta$,
$\beta_3\neq 3\beta$, and so on, since the strong
coupling fixed point is non-Gaussian, but it remains to be seen whether
$\beta_m$ is indeed independent of $m$.
%If, however, all the $\beta_m$ are indeed equal,
%there is presumably a simple explanation.
%One possible reason could be that the probability distribution function (PDF)
%$P\{ n(x) \}$
%can be decomposed into an absorbing part represented by a delta function
%at the origin, and an active part with small amplitude proportional to
%$(\delta r)^{\beta}$. ****COMMENT ON 0d***
%However, precisely how such a scenario might occur in MNS
%is unclear.
It is interesting to note that for the zero-dimensional (single-variable)
case, where an exact solution is available\cite{Brand,Graham},
all the moments do scale
with exactly the same exponent: $<n^{m}>=2^{m/2}\Gamma
[\frac{m+r/D}{2}]/\Gamma(-r/2D)$, i.e., $\beta_{m}=1$ for all
$m$.

To determine the other critical exponents, we need to calculate the two-point
function $C(x,t)=<n(x+x_{0},t+t_{0})n(x_{0},t_{0})>$ in the active phase.
We have computed
both $C(x,0)$ and
$C(0,t)$, for
$-r=2.8,2.6,2.5,2.4,2.34,2.3,2.2,2.215$.  We used
N=512, and averaged over
$T=10^6 \Delta t$ time steps in steady state, and over
up to 100 independent runs.

We summarize the results in
fig. 2. In fig. 2(a,b), $C(x,0)$ and $C(0,t)$
are plotted for different values of $r$. It is clear from these
figures that the scaling regime becomes bigger
as we approach the critical point, and that the amplitudes of the correlation
functions become smaller, consistent with $C(x,t)$ vanishing, as it must, in
the absorbing phase.
In fig. 2(c), we plot the absolute value of
the {\it connected} space and time correlation function, $C_{c}(x,t)=
C(x,t)-M^2$,
at the largest value of $r$, viz., -2.215.
The power law decay of the correlation functions in the scaling regime
can be characterized
by $C_c (0,t)=A_{t}(r)t^{-\alpha_{t}}$ and
$C_c (x,0)=A_{x}(r)x^{-\alpha_{x}}$ with
the exponents $\alpha_{t}=1.08\pm 0.04$ and $\alpha_{x}=1.65\pm 0.07$.
In conventional notation, this corresponds to exponents $\eta \equiv 2-d+
\alpha_{x}=2.65\pm 0.07$, and $z \equiv
\alpha_{x}/\alpha_{t}=1.53\pm 0.07$. In
fig. 2(d), we plot the dependence of the amplitudes of the spatial and temporal correlation
functions, $A_{x}(r)$ and $A_{t}(r)$, on $\delta r$ in the scaling regime;
these can be
fitted by $A_{x,t}(r)\sim\delta r^{\Delta}$ with the
exponent $\Delta =1.7\pm 0.07$.

>From these measured values of the four independent exponents $\beta$, $\eta$,
$z$, and $\Delta$, we obtain the correlation
length exponent $\nu$ from the scaling relation\cite{legend}
$\nu=(2\beta-\Delta)/(d-2+\eta)=1.03\pm 0.05$.
These values of $\nu$ and $z$ are in excellent agreement with
the values, 1 and 3/2 respectively, following from
our argument that the critical behavior of MNS is controlled by the
KPZ equation; $\beta_{2}$ is also $>1$, consistent with the bound derived
from KPZ.
To further check the accuracy of various
scaling exponents, we have also measured the decay of the average
density right at the critical point, starting from a homogeneous
initial condition: $M\sim t^{-\theta}$. Using scaling arguments,
it is easy to express $\theta$ in terms of other exponents:
$\theta=\beta/(\nu z)$.  From the numerical values of $\beta$, $\nu$,
and z, we predict $\theta$ to be 1.079.  In fig. 3, we plot
M versus time at $r=r_c$, and the exponent $\theta$ thus measured is
$\theta=1.1\pm 0.05$, in excellent agreement with the scaling prediction.
In the same figure, we have
included the behavior of the higher order moments $M_{2,3,4}(t)$ at
the critical point.  These plots strongly suggest
that the exponents $\theta_m = \beta_{m} / (\nu z)$
for these higher moments are all equal to
$\theta$, in
agreement with the static measurement.

Another important characterization of any phase transition is the
response function. For equilibrium systems, the fluctuation-dissipation
theorem ensures that the response function is directly related to
the correlation function of the system, so the susceptibility
probes the system's internal fluctuations
and correlations directly. However, the fluctuation-dissipation theorem
does not generally hold for nonequilibrium systems, where the
response and correlation functions can differ significantly.
Indeed, one of the
striking features of MNS is that the susceptibility is
predicted\cite{legend} to diverge over
a finite range of control parameters, while
the correlation function has, more conventionally, a unique critical point.

%**********************************
%The line of divergent susceptibility is determined by
%the non-renormalizibility
%of the response function in the absorbing state, and it actually extends
%into the active state, as we have studied in our previous work and
%shown exactly in the zero dimensional case.
%**********************************

%In previous work, we argued that for any $d$, the susceptibility diverges
%in a region of the absorbing phase contiguous with the critical point.
%(where there is no renormalization
%of the response function due to nonlinearities),
To test this prediction numerically in 1d, we
add a constant source term $\phi$
on the RHS of Eq. (\ref{discr}),
and calculate the average density $M(\phi,r)$ for
small values of $\phi$ and for each value of $r$. In figure 4(a),
we show the dependence of $M(\phi,r)$ on $\phi$ for different values of $r$
around the critical point $r_c$.  We can fit the curves for small $\phi$
by
$M(\phi,r)=M(0,r)+B(r)\phi ^{\gamma(r)}$ for small $\phi$, where $M(0,r)$
is the order parameter in the absence of any external source, and B(r) is
a function of $r$.
The static susceptibility is defined as
$\chi \equiv \lim_{\phi\rightarrow 0}\partial M/\partial\phi
=\lim_{\phi\rightarrow 0} B(r)\gamma(r) \phi^{\gamma(r)-1}$, so the susceptibility
diverges when $\gamma(r)<1$.  According to fig. 4(b),
where we show the dependence on $r$ of
the susceptibility exponent, $\gamma (r) -1$,
$\chi$
diverges not just in the absorbing phase, but
in the whole region $3.2>r>-5.9$ surrounding the
critical point $r_c=-2.18$.
Both the continuous variation of this exponent with $r$, and the divergence
of $\chi$ in both the active and absorbing phases near $r_c$ also occur in the
exactly solvable 0d problem\cite{legend}.

Another quantitative prediction of our previous
paper\cite{legend} is that the region of diverging $\chi$
should terminate in the absorbing phase
when the bare mass vanishes, i.e., at $r=0$ in the Ito representation.
However, because of the shift
in $r$ due to our numerical algorithm, the lower boundary
of this region is shifted to $r>0$.
A clearer way
to test the prediction is therefore to make the Hopf-Cole transformation
$n=e^{h}$ first, and then simulate the equation for $h$, noting
that the external field will have the form $\phi e^{-h}$.  Because
the Hopf-Cole relation automatically guarantees that the $n$ field is positive
definite, there is no need to alter the
numerical algorithm, so $\gamma(r=0)$ should equal 1. We have verified this
in a simulation.

We thank A. Pikovsky, P. Grassberger, T. Hwa, S. Chen, and N.-Z. Cao
for helpful discussions,
and T. Hwa for bringing refs. \cite{Magic} to our attention.

\begin{figure}
\caption{(a)$r_c(N)$ vs. inverse system size 100/N;
(b)$<n^{m}>$ with $m=1,2,3$ vs. control parameter $\delta r\equiv
r_c-r$; dashed line with slope 1.7 gives the best fit to the data.}
\label{fig1}
\end{figure}
\begin{figure}
\caption{(a) and (b) show the spatial and temporal correlation functions
with different values of $r<r_c$;
(c) {\it connected} correlation functions at $r=-2.215$; dashed lines
are fits with slopes 1.08 and 1.65; (d) amplitudes of the correlation
functions vs. $\delta r$; dashed line has slope 1.7. }
\label{fig2}
\end{figure}
\begin{figure}
\caption{Power-law decay in time
of various moments of the order parameter
at the critical point $r=r_c$.}
\label{fig3}
\end{figure}
\begin{figure}
\caption{(a) Order parameter vs. external field at values of
$r$ near $r_c$; dashed lines are fits to $M(\phi,r)-
M(0,r)\sim \phi^{\gamma(r)}$, as explained in text; (b) susceptibility
exponent vs. r; susceptibility
diverges for a range of $r$ values.}
\label{fig4}
\end{figure}

\end{multicols}

\begin{thebibliography}{99}
\bibitem{rho} Our simulations for $\rho=0$ produced strong
coupling exponents indistinguishable from those for $\rho=1$.
\bibitem{inverse} Standard finite-size scaling
predicts $r_c \sim
N^{-1/\nu}$, so $r_c$ scaling like $1/N$ is consistent with $\nu =1$.
\bibitem{discrete} Numerical calculations for 1d coupled map models believed
to be in the same universality class as Eq. (\ref{MNS})
also give a $\beta$ close to 1.70; see ref. 8b.
\end{thebibliography}
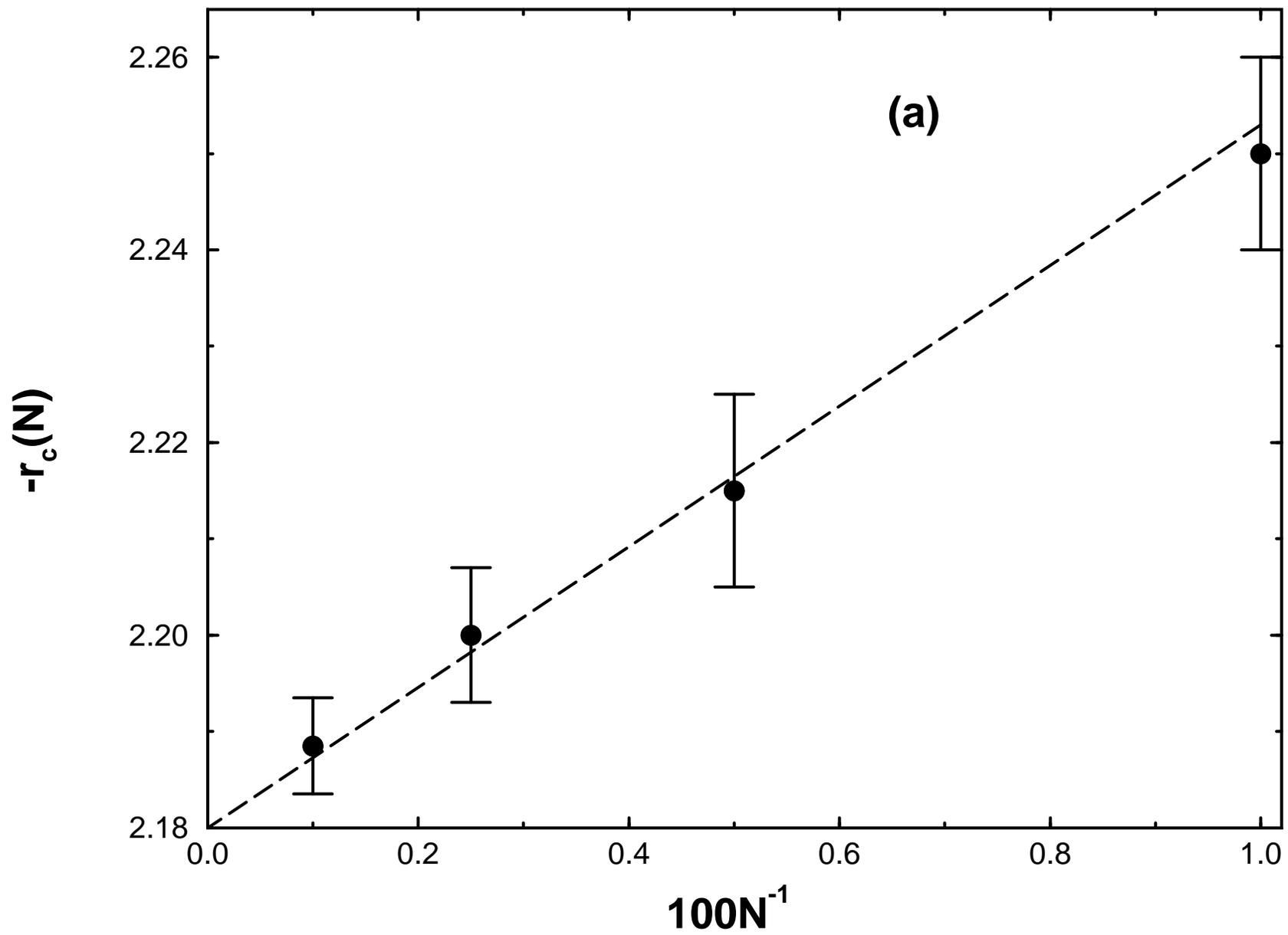
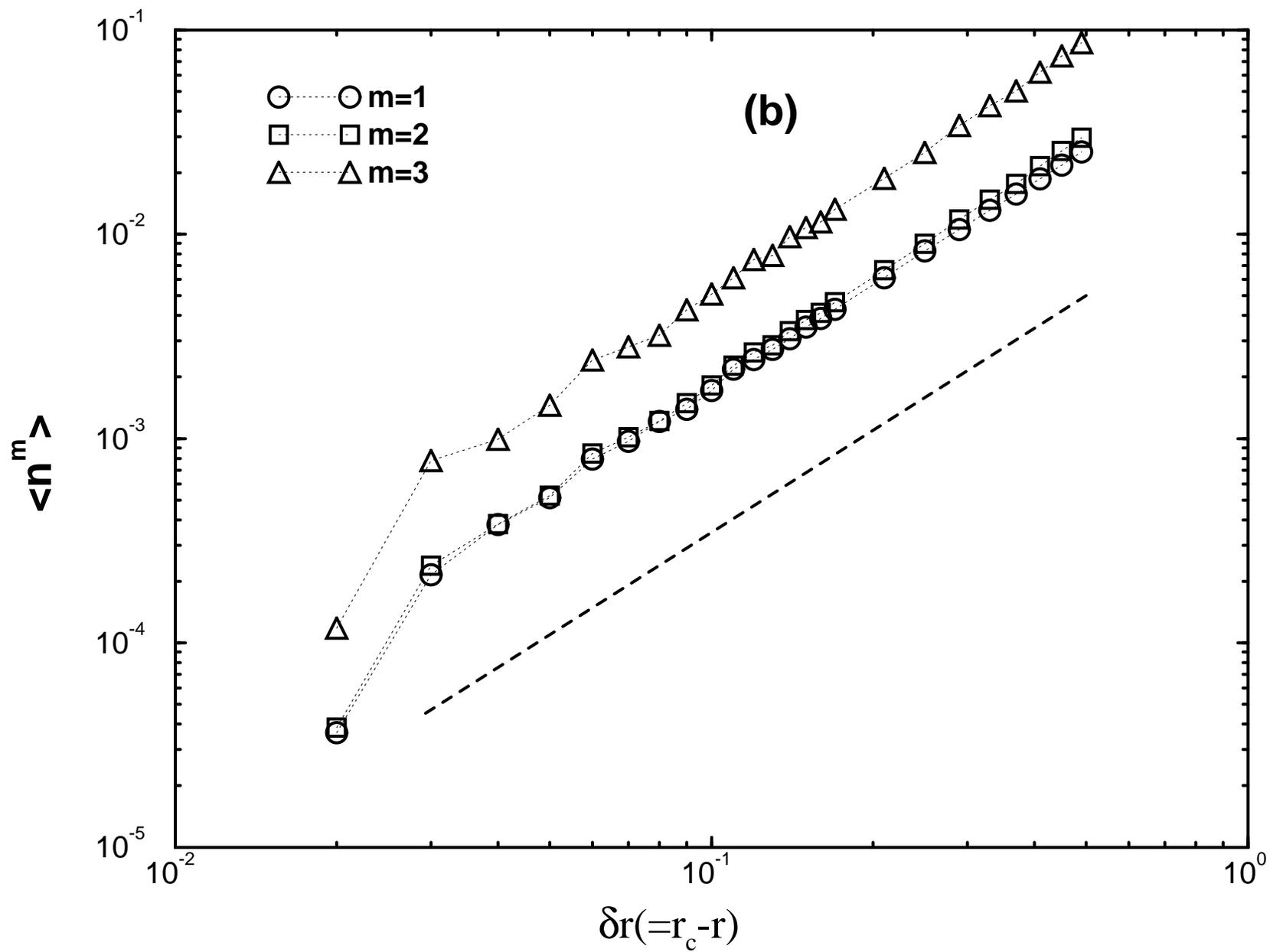
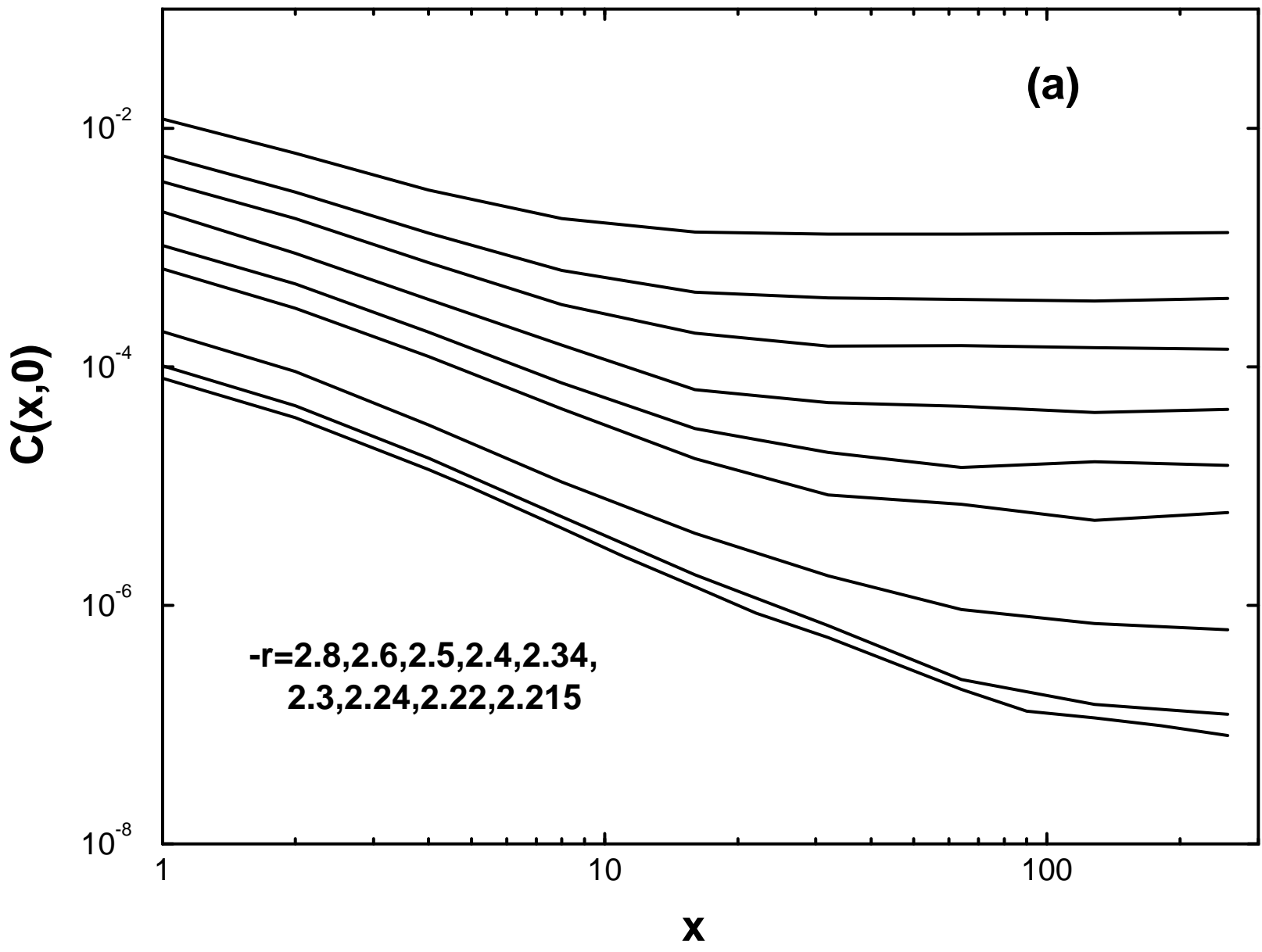
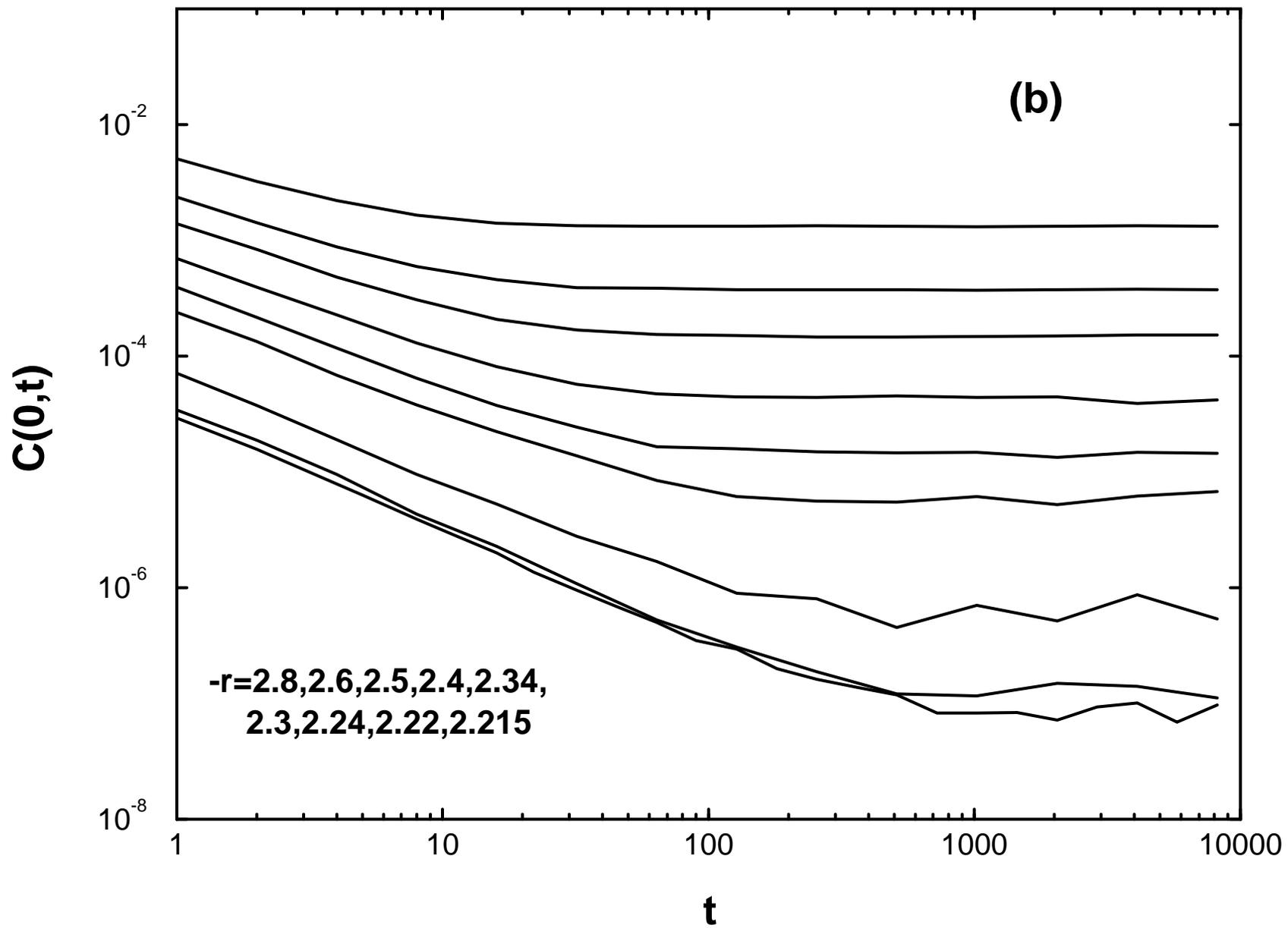
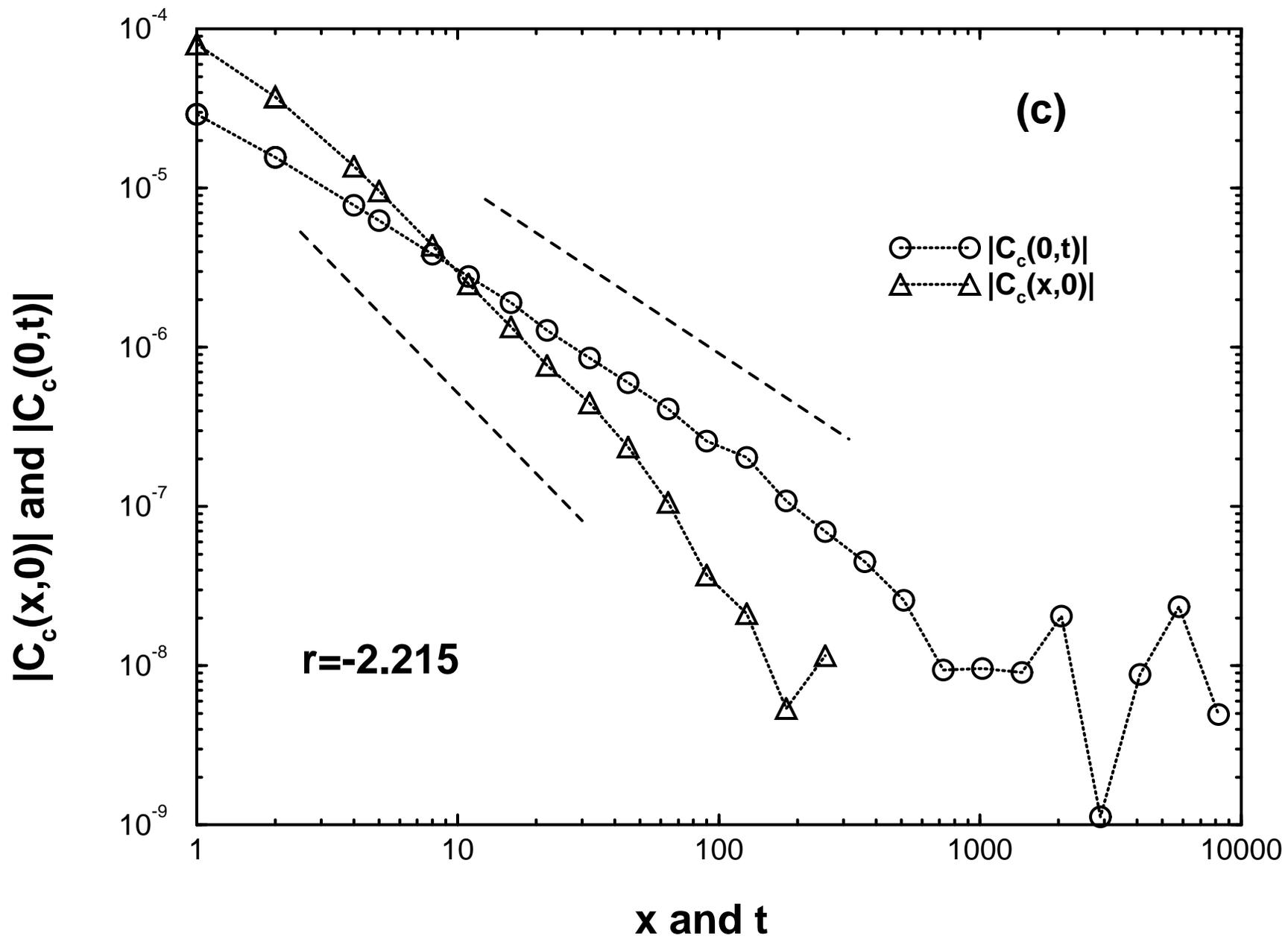
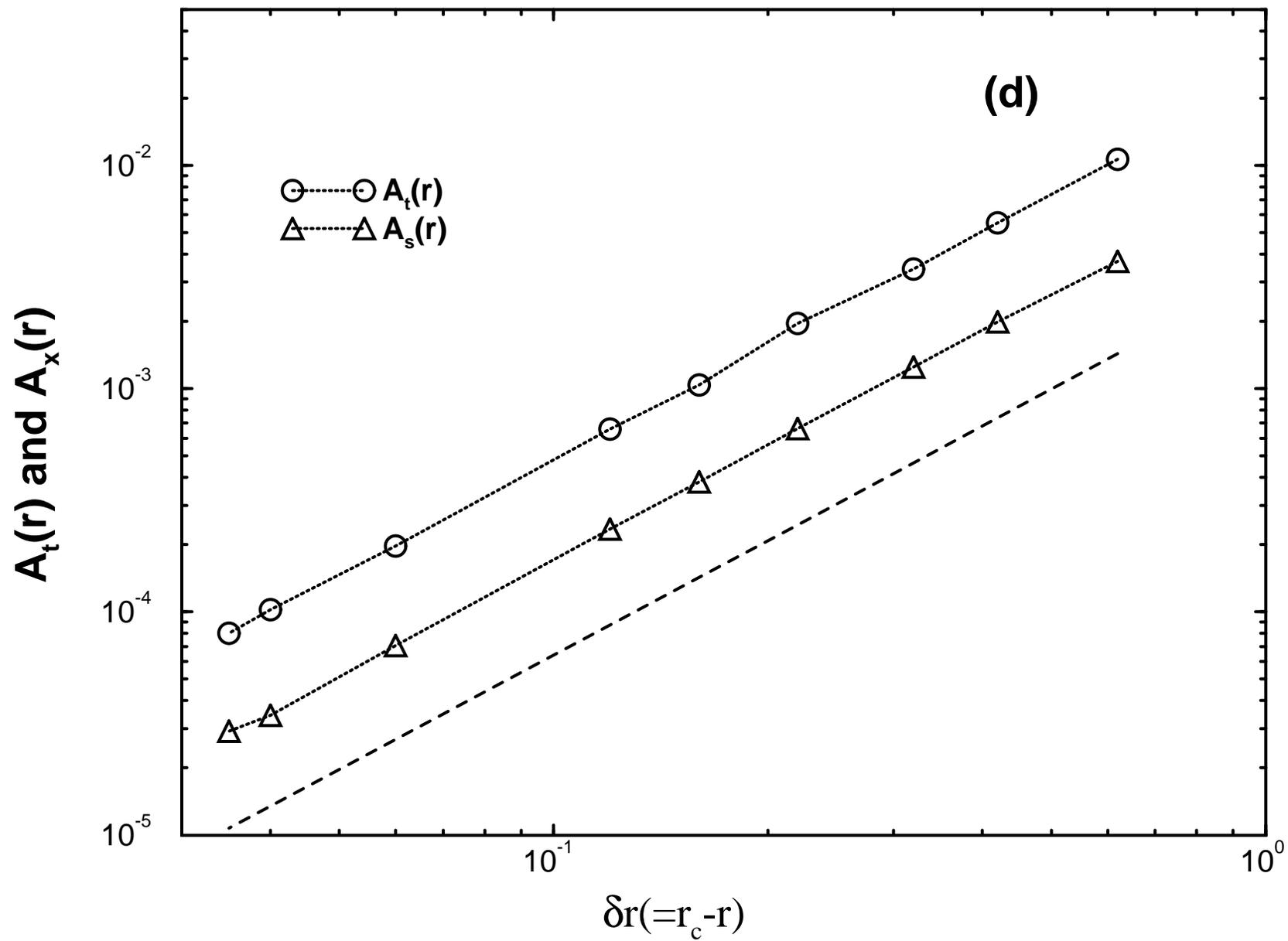
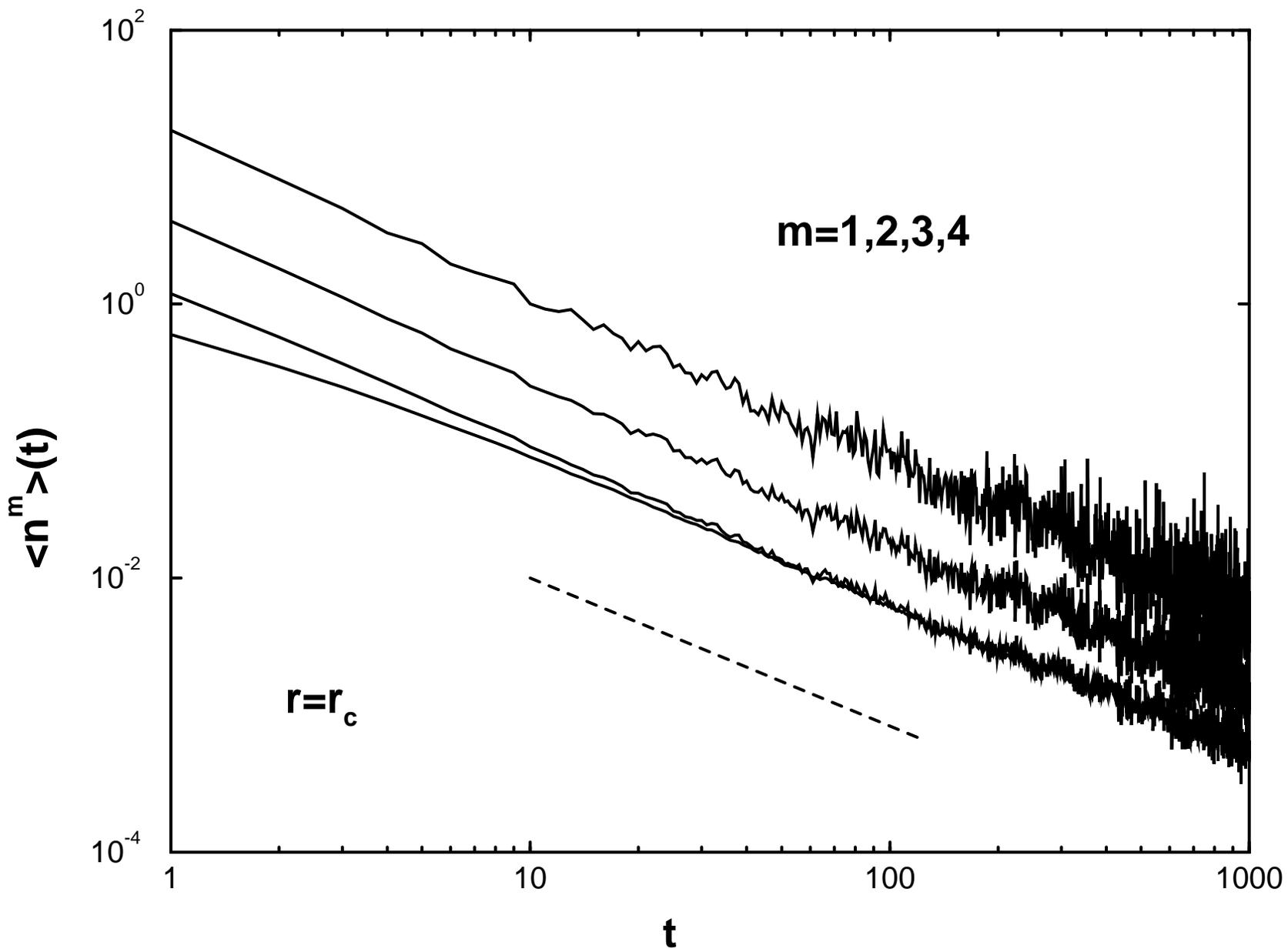
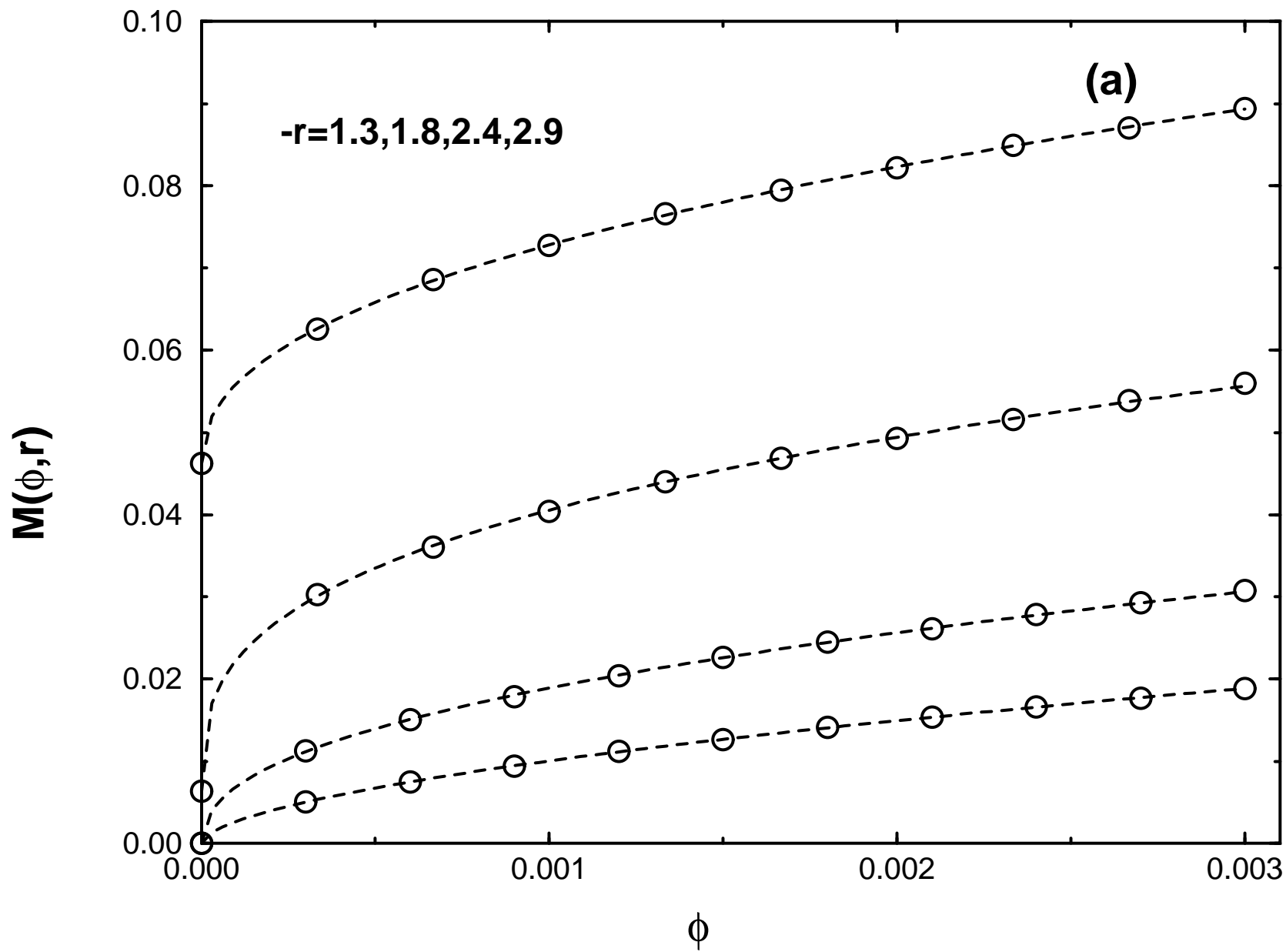
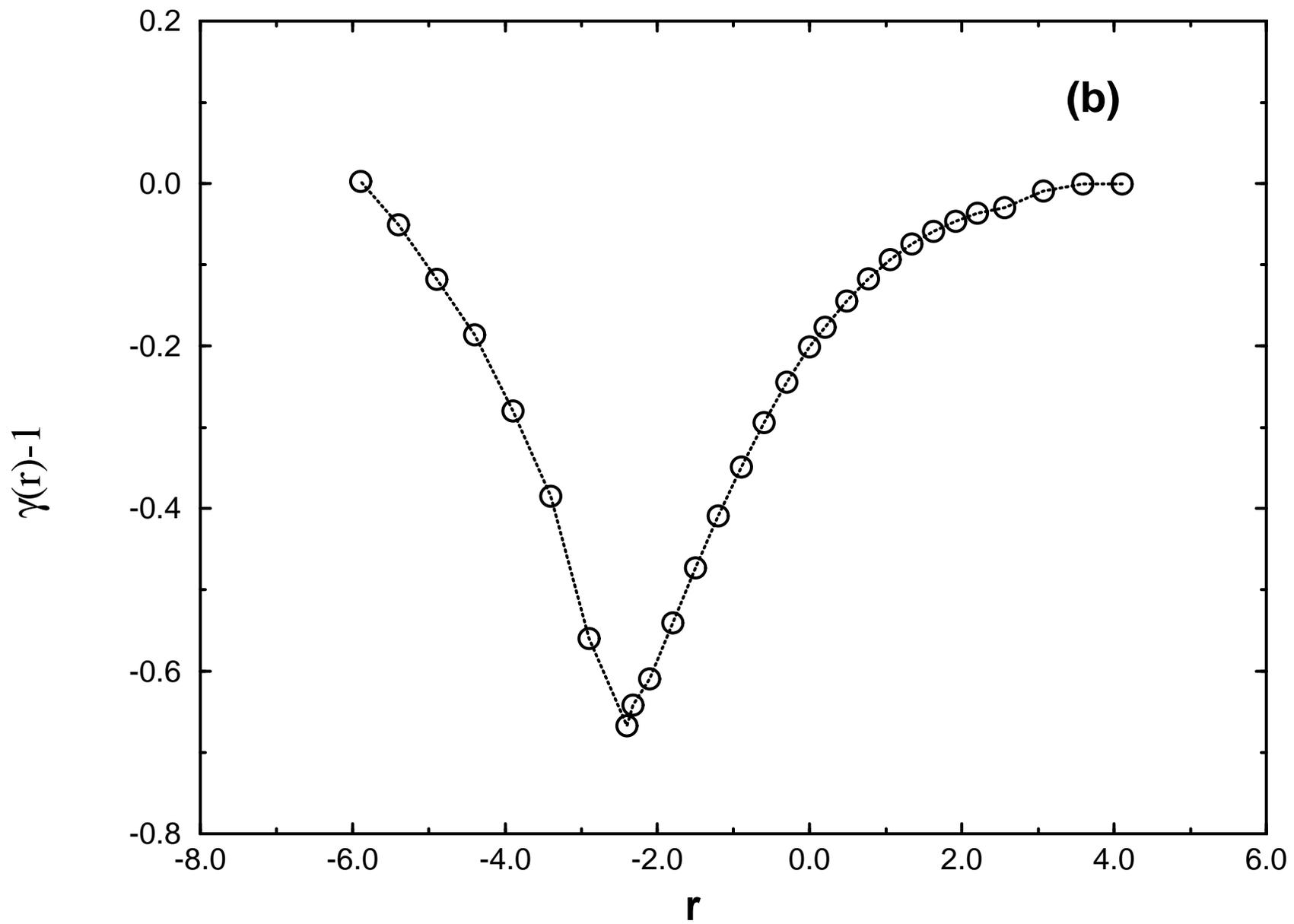
\end{document}